\documentclass{intech}

\booktitle{Will-be-set-by-IN-TECH}%

\chaptertitle{The s-process nucleosynthesis in massive stars: current status and uncertainties due to convective overshooting} 

\authors{M.~L. Pumo}
\affiliation{Universit\`a degli studi di Padova, Dip. di Fisica e Astronomia ``G. Galilei'', Vicolo dell'Osservatorio 3, I-35122 Padova\\
INAF-Osservatorio Astronomico di Padova, Vicolo dell'Osservatorio 5, I-35122 Padova\\
Bonino-Pulejo Foundation, Via Uberto Bonino 15/C, I-98124 Messina\\
INAF-Osservatorio Astrofisico di Catania, via S. Sofia 78, I-95123 Catania
}
\country{Italy}

\def\Msun{$M_\odot$}

\begin{document}

\maketitle

\section{Introduction}
From time out of mind, men wish to know more about the world surrounding them and, in particular, to understand the origin of 
the matter. However, only in the late 1950s, thanks to the pioneering work of \citet[][the famous B$^2$FH]{B2fh} and the independent analysis of \citet[][]{cameron57}, the basic principles of explaining the origin of the elements were laid down in the theory of nucleosynthesis, giving rise to a dedicated, interdisciplinary field of research referred to as ``nuclear astrophysics''.
  
Since the early days of its development, nuclear astrophysics has improved at an impressive pace. Nowadays, it is a peculiar mixing of knowledge that blends the progresses in experimental and theoretical nuclear physics, ground-based and space observational astronomy, cosmochemistry, and theoretical astrophysics.

Different works have been devoted to review the achievements reached by the nuclear astrophysics, either providing an overall picture of 
this field of research \citep[e.g.][]{Wallerstein97,AT99,JI11,RP11}, or focusing on various its subfields such as the Big-Bang nucleosynthesis and primordial abundances \citep[e.g.][]{Tytler00,Steigman07,Iocco09}, the nucleosynthesis mechanisms of specific types of trans-iron elements \citep[e.g.][]{Meyer94,Kappeler99,AG03,Arnould07,Kappeler11}, the hydrostatic and explosive stellar nucleosynthesis  \citep[e.g.][and references therein]{Busso99,Chiosi07,JH07,woosley2002}, the nucleosynthesis by spallation \citep[e.g.][]{Reeves94,Vangioni-Flam00}, the experimental techniques and theoretical methods used to investigate nuclear processes of astrophysical interest \citep[e.g.][]{Angulo09,Costantini09,TB03,Baur03,BG10,Spitaleri10,Thielemann01}. 

This chapter is concerned with one of the aforementioned subfields of the nuclear astrophysics. In particular, it is devoted to 
the so-called s-process, which is a nucleosynthesis mechanism responsible for the production of about half of all the trans-iron elements. The chapter starts out in Sect.~\ref{sec:abundances} with some basic considerations on the solar system composition, considered as ``standard of reference'' dataset for cosmic abundances. A brief description of the nucleosynthesis mechanisms responsible for the production of trans-iron nuclei follows in Sect.~\ref{sec:nucleosynthesis}. Afterward (see Sect.~\ref{sec:spro}), the s-process nucleosynthesis mechanism is reviewed and its different components are discussed. A specific attention is paid to the so-called ``weak component'' occurring in massive stars ($M_{ZAMS} \gtrsim 13 M_{\odot}$), showing its sensitivity to stellar mass and metallicity (see Sect.~\ref{sec:sensitivity}). Moreover the uncertainties affecting the efficiency of this component are described (see Sect.~\ref{sec:uncertainties}), placing particular emphasis upon uncertainties due to convective overshooting (see Sect.~\ref{sec:overshooting}). Prospects of improvements in modeling the s-process weak component and their possible consequences for some open astrophysical questions are also briefly discussed (see Sect.~\ref{sec:remarks}).


\section{General considerations on the origin of the elements}
\subsection{Cosmic abundances}
\label{sec:abundances}
Whatever nucleosynthesis model is built, a comparison between the model predictions to the observed cosmic abundances is needed. 

Even if the possibility of defining a truly ``standard'' set of observed cosmic abundances has to be considered with caution \citep[see e.g.][]{grevesse96}, the composition of the material from which the solar system formed $\sim$ 5.6 Gy ago (referred hereafter also as solar composition) is usually considered as the ``standard of reference'' dataset \citep[e.g.][and references therein]{asplund09}. Such a choice is essentially due to the fact that the solar composition is the only comprehensive sample with a well defined isotopic abundance distribution \citep[e.g.][]{Arnett96,Kappeler99}, since it can be derived using different sources of information that include, among others, the Earth, the Moon, other solar system planets, the Sun, meteorites, and material from the interplanetary medium. The methods employed to gather abundances information combine spectral analysis (e.g.~via spectroscopy of the solar photosphere), laboratory measurements of matter samples (e.g.~via mass spectrometry of meteorites, Lunar glasses, and material from the Earth's crust and carried by space probes from the interplanetary medium), and particle detection from space-based experiments (e.g.~via analysis of solar wind and solar energetic particles). 

\begin{figure}[ht]
\centering
\includegraphics[angle=-90,width=6.2cm]{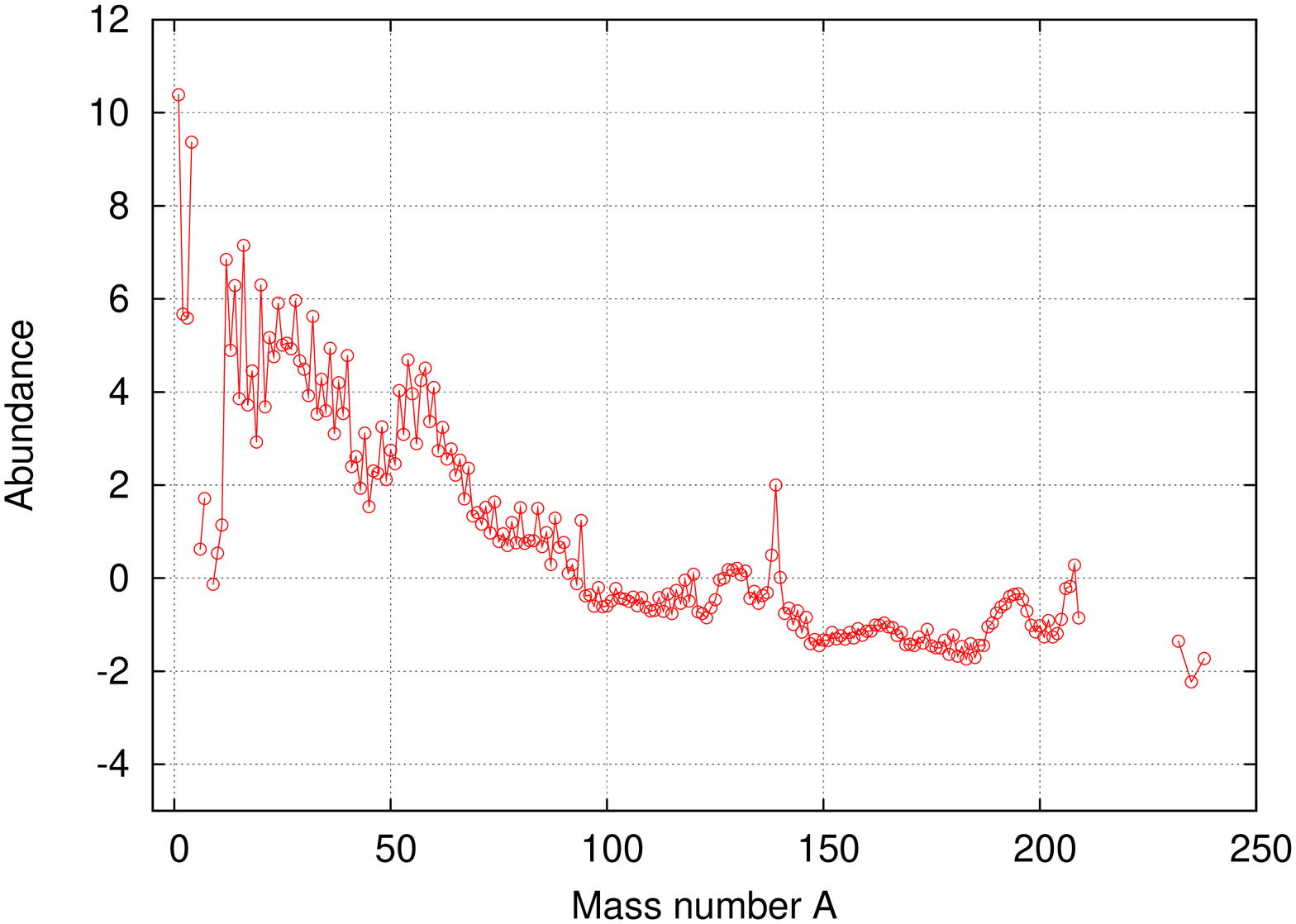}
\includegraphics[angle=-90,width=6.2cm]{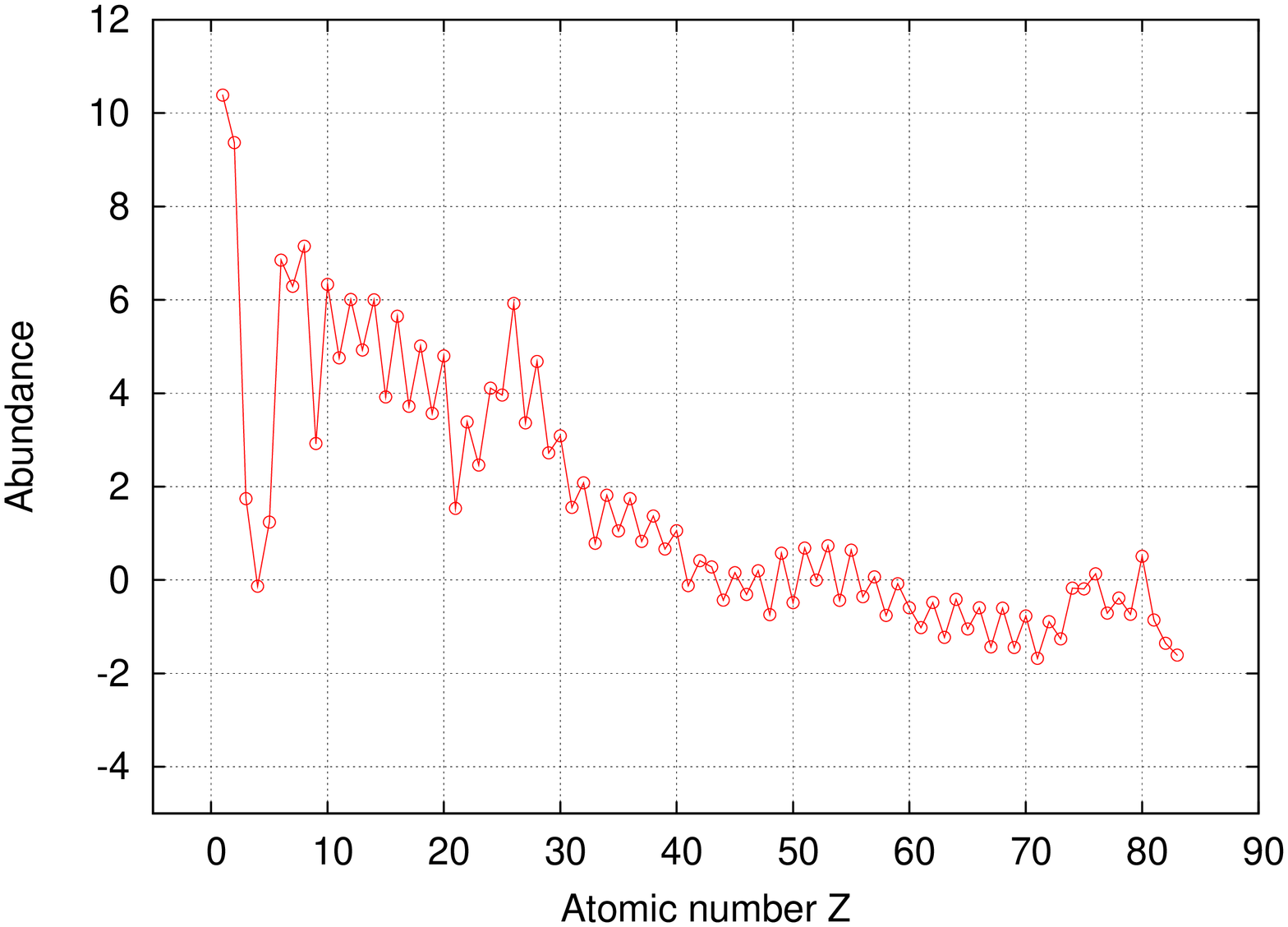}
\caption{Nuclear (left panel) and elemental (right panel) solar compositions from the compilation of \citet[][]{lodders03}. In both panels, the abundances are normalized to 10$^6$ Si atoms and the $y$ axes are logarithmic. In left panel, only the most abundant isobar is reported for a given mass number A.}\label{fig:abundances}
\end{figure}

Specifically, the elemental solar composition (displayed in right panel of Fig.~\ref{fig:abundances}) is largely grounded on abundance analyses of a peculiar class of uncommon meteorites --- the so-called CI chondrites --- which are believed to reflect the composition of the ``primitive'' solar system. However information derived from the Sun (i.e.~using the solar photospheric spectrum, the impulse flare spectra, and the analysis of solar wind and solar energetic particles) have to be used for determining the abundances of H, C, N, O and the noble gases He, Ne, and Ar; while analyses based on theoretical considerations are required for the ``heavy'' noble gases Kr and Xe. The nuclear solar composition (displayed in left panel of Fig.~\ref{fig:abundances}) is determined considering the terrestrial isotopic compositions as the most representative ones for all the elements of the primitive solar system with the exception of the H and all the noble gases, for which other sources of information (e.g.~Jupiter's atmosphere, solar wind and lunar samples) are used (for details see e.g.~\citeauthor{lodders03}, \citeyear{lodders03}, but also the other widely used compilations of the solar composition by \citeauthor{AA89}, \citeyear{AA89}; \citeauthor{GN93}, \citeyear{GN93};  \citeauthor{GS98}, \citeyear{GS98}; \citeauthor{lodders09}, \citeyear{lodders09}, and \citeauthor{asplund09}, \citeyear{asplund09}).
 
 
Looking at Fig.~\ref{fig:abundances}, it is easy to notice that the solar distribution is clearly dominated by the H and He abundances, followed by a ``deep'' (with respect to the neighboring nuclides) minimum of the abundances for the elements Li, Be, and B. Subsequently, the main feature of the distribution is the presence of a series of peaks at the locations of the ``$\alpha$-elements\footnote{Alpha-elements are so-called since their most abundant isotopes are integer multiples of the mass of the $\alpha$ particle. The most abundant are $^{16}$O and $^{12}$C, followed by $^{20}$Ne, $^{24}$Mg, $^{28}$Si, $^{32}$S, $^{36}$Ar, and $^{40}$Ca.}'', superimposed on an exponentially decreasing curve from the A $\simeq$ 12-16 mass region down to the Sc, followed by a pronounced peak centered around $^{56}$Fe. From this peak on, the distribution becomes fairly flat with a variety of superimposed peaks which correspond to nuclides having magic numbers\footnote{In the nuclear shell model, a magic number is a number of nucleons (either protons or neutrons) such that they can be arranged into complete shells within a nucleus. The seven most widely recognised magic numbers are: 2, 8, 20, 28, 50, 82, and 126. Nuclei consisting of such a magic number of nucleons have a particularly tightly bound configuration \citep[see e.g.][for details]{Povh08}.} of neutrons.


As already pointed out by \citet[][]{B2fh}, these features are a ``reflection'' of the different nucleosynthesis processes responsible for the production of the various isotopes. Even without going into details because it is out of the purpose of this chapter \citep[for details the interested reader is referred to the reviews/books of][and references therein]{Arnett96,AT99,Clay,Kappeler99,JI11,RR88}, we remind that the nuclides with A $<$ 12 are produced by Big-Bang nucleosynthesis and via spallation mechanisms. The minimum at the elements Li, Be, and B primarily reflects the difficult to synthesize such rare and fragile nuclides due to the stability gaps at A~$=$~5 and 8. Various charged-particle induced reactions, which occur inside the stars during quiescent evolutionary phases and are accountable for the stellar energy production, are instead responsible for the nucleosynthesis of all the elements from the A $\simeq$ 12-16 mass region up to the region of the iron peak. The exponentially drop in the elemental abundances up to the Sc is just related to the increase of the Coulomb barrier with increasing the nuclear charged of the particles involved in the different charged-particle induced reactions, that increasingly hampers such reactions. Exceptions to this trend are the peaks at the locations of the $\alpha$-elements and the iron peak. The greater stability of these nuclei (compared to the one of the neighbouring nuclides) leads to a their more abundant production and, consequently, gives rise to the above mentioned local maxima in the solar distribution. Finally, neutron capture chains and an additional mechanism characterized by photo-disintegrations of preexisting nuclei are called for in order to synthesize the trans-iron elements, as better described in the following Sect.~\ref{sec:nucleosynthesis}. Indeed, if the trans-iron elements had been synthesized by charged-particle induced reactions, the peaks present in the trans-iron region of the solar distribution would not be explained and the abundances of the trans-iron elements would exhibit a much higher decrease (with the mass number A) than the one actually observed in the solar distribution.  


\subsection{Nucleosynthesis mechanisms for the production of trans-iron nuclei}
\label{sec:nucleosynthesis}
\begin{figure}[ht]
\centering
\includegraphics[angle=+0.52,width=11.5cm]{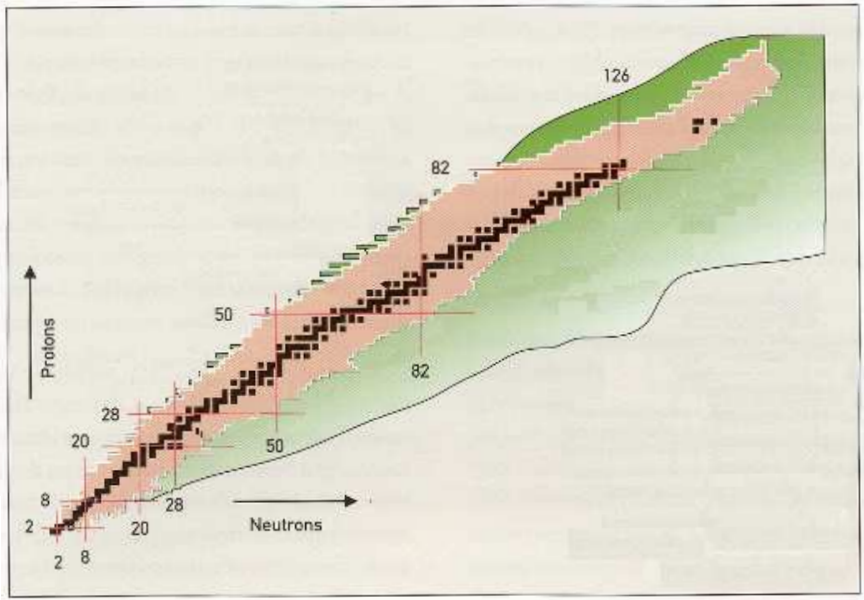}
\caption{Chart of nuclides where bound nuclear systems are reported as a function of the neutron number N ($x$ axis) and the proton number Z ($y$ axis). The brown squares represent stable plus very long-lived (lifetime $>$ 10$^5$ yr) nuclei that define the so-called valley of nuclear stability. The red lines show the magic numbers for the valley of nuclear stability. The pink region designates the zone of unstable nuclei that has been explored in laboratory; while the green region indicates the so-called ``terra incognita'' (Latin expression for ``unknown land'') occupied with unstable nuclei that remain to be explored. The explorable nuclear landscape is bounded by the proton drip line to the upper left and neutron drip line to the bottom right. Both drip lines are marked by thin lines. (Figure adapted from http://www.pas.rochester.edu/$\sim$cline/Research/sciencehome.htm)
\label{fig:valley}}
\end{figure}

The stable trans-iron nuclides can be classified into three categories according to their position on the chart of nuclides (see Fig.~\ref{fig:valley}): those located at the bottom of the valley of nuclear stability, called the s-nuclei, and those situated on the neutron-deficient or neutron-rich side of the valley, named the p- or r-nuclei, respectively. 

As pointed out above, the charged-particle induced reactions are not able to account for the production of these three types of trans-iron nuclides, which are instead primarily produced by three different mechanisms, naturally referred to as the s-, r-, and p-processes.

The first two processes (namely, s- and r-processes) can take place through neutron captures and subsequent $\beta$-decays in astrophysical environments where at least one neutrons source is working efficiently \citep[e.g.][]{Kappeler99}. As a result of each (n,~$\gamma$) capture reaction, a generic nucleus (Z,~A) is transformed into the heavier isotope (Z,~A+1). If this isotope is stable against $\beta$-decay, an additional neutron capture can take place, leading to the isotope (Z,~A+2). Otherwise, if the produced isotope is unstable, it can either decay into the isobar (Z+1,~A+1) or capture another neutron \citep[e.g.][]{Clay}. The question whether this unstable isotope decays or captures a neutron depends on the values of $\tau_{\beta}$ and $\tau_{n\gamma}$, which are the $\beta$-decay lifetime and the average time between two successive neutron captures for such unstable isotope, respectively. If the relation $\tau_{\beta} \gg \tau_{n\gamma}$ is valid for the majority of the unstable nuclides involved in the nucleosynthesis process, the sequence of neutron captures and $\beta$-decays is called s-process (``s'' stands just for ``slow'' neutron capture); otherwise, if neutron capture proceeds on a rapid time scale compared to the $\beta$-decay lifetimes (i.e. relation $\tau_{\beta} \ll \tau_{n\gamma}$ is valid for the majority of the unstable nuclides), the sequence of reactions is named r-process, where ``r'' stands for ``rapid'' neutron capture \citep[e.g.][]{RR88}.

\begin{figure}[ht]
\centering
\includegraphics[angle=+0.47,width=13.0cm]{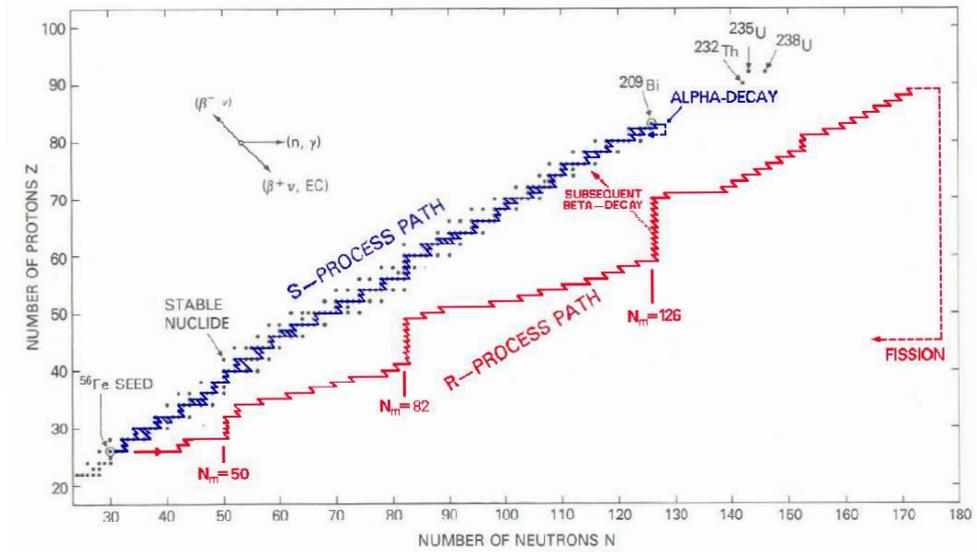}
\caption{Paths of the s- and r-processes in the N-Z plan. Both processes start with the nuclides of the iron peak region (mainly $^{56}$Fe) as seeds. The s-process path (blue line) follows the valley of nuclear stability and stops at the $^{209}$Bi, after which one enters the region of $\alpha$-instability. The r-process path (red line) moves far to the neutron-rich side of the valley of nuclear stability, bypasses nuclei with natural $\alpha$-radioactivity (which stops the s-process path), and terminates by $\beta$-delayed fission and neutron-induced fission at A~$\simeq$~270 \citep[e.g.][]{Thielemann01}. The chains of $\beta$-decays from the path of the r-process towards the valley of nuclear stability, which occur after the r-process neutron irradiation, are schematized by a dashed line. The sequences of waiting points at nuclei with magic neutron numbers N$_m$ = 50, 82, and 126 are also highlighted. \citep[Figure adapted from][]{RR88}
\label{fig:srpro}}
\end{figure}

Fig.~\ref{fig:srpro} reveals that the s-process involves the addition of neutrons to seed nuclei, which are nuclear species of the iron peak region (mainly $^{56}$Fe). In particular, this nucleosynthesis process produces nuclei from the A $\simeq$ 60 mass region up to $^{209}$Bi, closely following the valley of nuclear stability. Indeed, since the neutron capture is slow compared to the $\beta$-decay rates, the neutron capture chains move through the stable isotopes of a given element until an unstable isotope is reached. At this point, a $\beta$-decay occurs and the neutron capture chains resume in the element having the nuclear charge increased by one unit \citep[e.g.][for details]{Clay}.

As with the s-process, also the r-process has the nuclides of the iron peak region as seeds. However, the path of the r-process moves along the extreme neutron-rich side of the valley of nuclear stability (see Fig.~\ref{fig:srpro}), where the values of the neutron binding energies approach zero (so-called neutron drip line; see also caption of Fig.~\ref{fig:valley}). Indeed, since the neutron capture proceeds on very rapid time scale compared to the $\beta$-decay lifetimes, a generic nucleus (Z,~A) is transformed into the heavier neutron-rich isotope (Z,~A+i) by a series of (n,~$\gamma$) capture reactions. Only when it is reached a point where the (n,~$\gamma$) capture reaction and its inverse ($\gamma$,~n) reaction are in equilibrium, capture reactions stop and a $\beta$-decay can occur, transforming the nucleus (Z,~A+i) into its isobaric neighbour (Z+1,~A+i). A this point, the chain of capture reactions restarts and the nucleus (Z+1,~A+i) absorbs neutrons until the balance between capture reactions and photodisintegrations is again reached for its isotopic neighbour (Z+1,~A+i+k). This recurring sequence of ``neutron captures plus $\beta$-decay'' reactions where it is necessary to ``wait'' a $\beta$-decay prior of restarting another series of neutron captures leads to a so-called waiting point on the N-Z plan. In particular, if the r-process path reaches a nucleus with a magic neutron number N$_m$, the next heavier isotope with N$_m$+1 neutrons has a relatively small neutron binding energy and it is relatively easy to reach the equilibrium between the (n,~$\gamma$) capture reaction and its inverse ($\gamma$,~n). Consequently, the sequence of neutron capture events stops and, after one neutron capture and a subsequent $\beta$-decay, the (Z,~N$_m$) nucleus is transformed into its isobaric neighbour (Z+1,~N$_m$), which is again a nucleus with the same magic number N$_m$ of neutrons. Therefore, it is expected the occurrence of a series of waiting points at the same magic neutron number N$_m$, that increases the Z value by one unit at a time and moves the resulting nuclei nearer and nearer to the valley of nuclear stability. Only when these nuclei are enough close to the valley of nuclear stability that the neutron binding energy becomes sufficiently large to break through the neutron magic bottleneck at N$_m$, the chain of capture reactions restarts (e.g.~see in Fig.~\ref{fig:srpro} the sequences of waiting points at nuclei with N$_m$~=~50, 82, and 126). After the synthesizing event (when the neutron irradiation stops), the very neutron-rich unstable nuclides undergo chains of $\beta$-decay, which end at the most neutron-rich stable isobar for each value of A. In this way, it is possible to produce either neutron-rich stable nuclei bypassed by the s-process\footnote{Such neutron-rich stable nuclei are referred to as ``r-only`` products. Similarly, nuclides produced by the s-process and bypassed by the r-process are referred to as ``s-only'' nuclei.} or nuclides lying also on the s-process path \citep[see e.g.][for details]{RR88,Clay,Arnould07,AT99}.  

The p-process is instead responsible for the production of proton-rich isotopic species (named p-nuclei), which are skipped by the s-process and r-process paths because they are ``shielded'' from formation both via $\beta$-decay by the presence of other stable isobars and via neutron capture by the lack of other less massive stable isotopes (neutron capture will obviously never bring a more massive stable isotope to a lighter one). The general consensus is that the p-nuclei are synthesized through ``photo-erosion'' of neutrons, $\alpha$ particles and protons involving heavy ($A\gtrsim 75$) isotopes previously formed via the s- and/or r-processes. A variety of astrophysical sites (e.g.~the Ne-O-rich layer of massive stars during their pre-supernova phase or their explosion as core-collapse supernova, the C-rich zones of Chandrasekhar-mass white dwarfs exploding as Type Ia supernovae, and the exploding sub-Chandrasekhar mass white dwarfs on which He-rich material has been accreted from a companion star) have been identified as possible active p-nuclei contributors, but a complete and self-consistent model for the origin of the p-isotopes remain to be soundly based \citep[for details see e.g.][and references therein]{AG03}.

\section{The s-process}
\label{sec:spro}

\subsection{An analytical approach: the $\sigma$N-curve and the different components of s-process}
\label{sec:analytical_approach}
As mentioned above, the path of the s-process follows the valley of nuclear stability, synthesizing about half of all the trans-iron elements. The time dependence of the abundance of these synthesized nuclei can be described on first approximation\footnote{A dedicated numerical modelling is required for a more detailed description \citep[see e.g.~the numerical approaches described in][and references therein]{prantzos87,pignatari08,pumo10}.} using an analytical approach. It is based on the assumption that the relation $\tau_{\beta} \gg \tau_{n\gamma}$ is valid for all the unstable nuclides involved in the nucleosynthesis process and, consequently, that a generic unstable nucleus (Z,~A+1) immediately decays into its isobar (Z+1,~A+1), which is usually stable. Although this assumption does not hold in every instance (cf.~Sect.~\ref{sec:branchings}), it enable us to describe the s-process in a relatively simple way and with no general damage to the theoretical description of this nucleosynthesis process \citep[e.g.][for details]{Clay,RR88}. Indeed, according to such assumption, one can neglect the abundances of unstable species and assume that, fixed the mass number $A$, there is only one stable nuclide with a given $Z$ (i.e.~only one isotope for each element) involved in the nucleosynthesis process. As a consequence, the time dependence of the abundance $N_A$\footnote{The abundance is labeled only by the mass number $A$ and not additionally by the nuclear charge $Z$ because the latter quantity is uniquely defined within the aforementioned assumption that there is only one stable nuclide with a given $Z$ involved in the nucleosynthesis process.} of the nuclei synthesized by the s-process can be written as
\begin{equation}
  \frac{dN_A(t)}{dt}=N_n(t)N_{A-1}(t)<\sigma v>_{A-1}\,-\,N_n(t)N_A(t)<\sigma v>_A, 
  \label{eq:fond_pro_s}
\end{equation}
where $N_n(t)$ is the neutron density at time $t$, while $<\sigma v>_{A-1}$ and $<\sigma v>_{A}$ are the reaction rate of the capture reaction involving the isotope with mass number $A$~$-$~$1$ and $A$, respectively. The first term on the right side of the above equation describes the production of an isotope with mass number $A$ by neutron capture of its lighter ``neighbour''  with mass number $A$~$-$~$1$, and the second one represents the destruction of the isotope with mass number $A$, again due to a neutron capture reaction. All the terms in the equation are time-dependent because of their dependence upon the temperature that, in turn, depends on the time. 


Relation \ref{eq:fond_pro_s} defines a set of coupled differential equations, which can be solved analytically under the additional assumption that the temperature is essentially constant during the event of neutron irradiation. Indeed, with this further hypothesis, it is possible to replace $<\sigma v>_{A-1}$ with $\sigma_{A-1}\,v_T$, where $\sigma_{A-1}$ is the Maxwellian-averaged neutron-capture cross section for the isotope with mass number $A$~$-$~$1$ and $v_T$ is the thermal velocity given by the relation $v_T$~$\simeq$~$\left(\frac{2kT}{M_n}\right)^{1/2}$, in which $k$ and $T$ have their standard meaning of Boltzmann constant and temperature, and $M_n$ is the neutron mass \citep[see e.g.][for details]{RR88}. Similarly $<\sigma v>_{A}$ can be replaced with $\sigma_{A}\,v_T$ and, consequently, relation \ref{eq:fond_pro_s} becomes
\begin{equation}
  \frac{dN_A(t)}{dt}=v_T\,N_n(t)\left[N_{A-1}(t)\sigma_{A-1}\,-\,N_A(t)\sigma_A\right]. 
  \label{eq:fond_2}
\end{equation}
Moreover it is possible to introduce a new variable, the so-called time-integrated neutron flux $\tau$, which is a measurement of the total neutron irradiation per unit of area and is defined as $\tau=v_{\,T}\int_0^tN_n(t)dt$. Replacing the time $t$ with this new variable $\tau$, the relation \ref{eq:fond_2} reduces to
\begin{equation}
  \frac{dN_A(t)}{d\tau}=\sigma_{A-1}N_{A-1}\,-\,\sigma_{A}N_A, 
  \label{eq:fond_3}
\end{equation}
where the rate of change of the $N_A$ is now with respect to $\tau$.

\begin{figure}[ht]
\centering
\includegraphics[width=11.0cm]{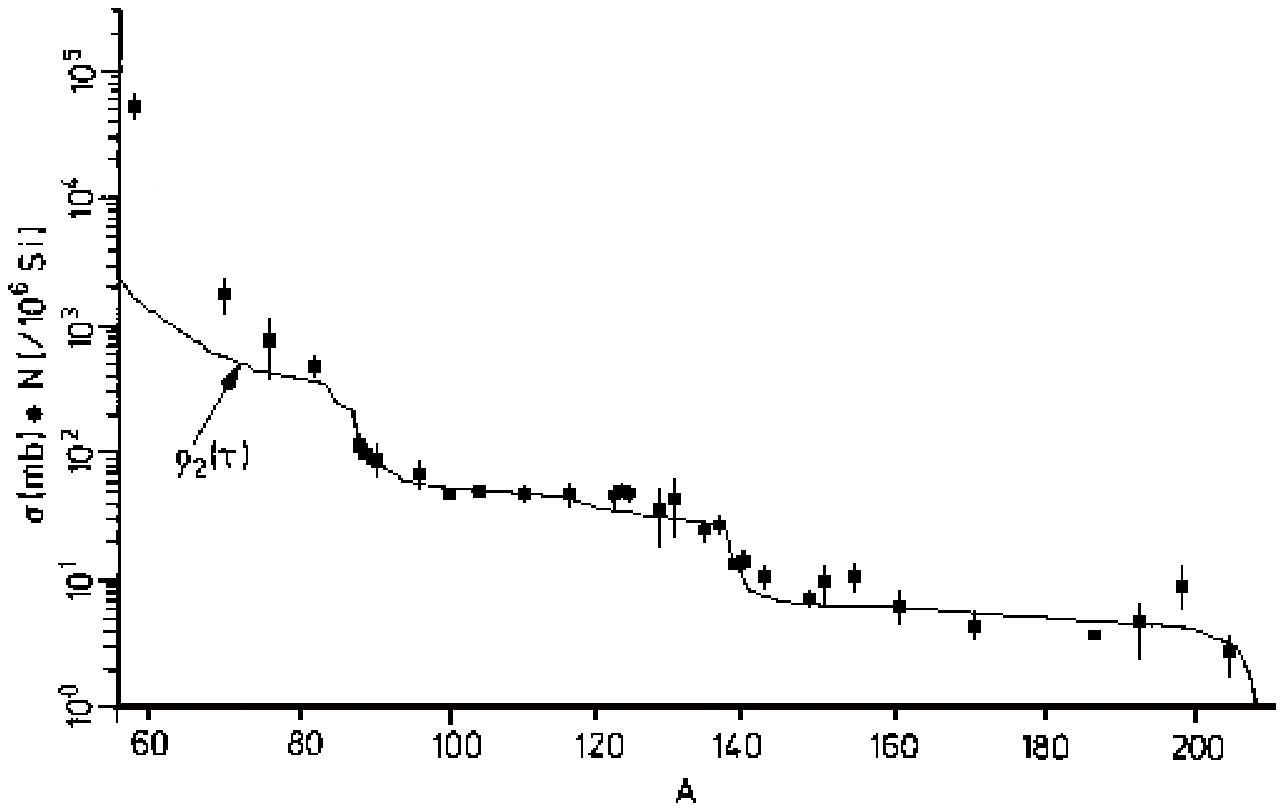}
\includegraphics[width=11.0cm]{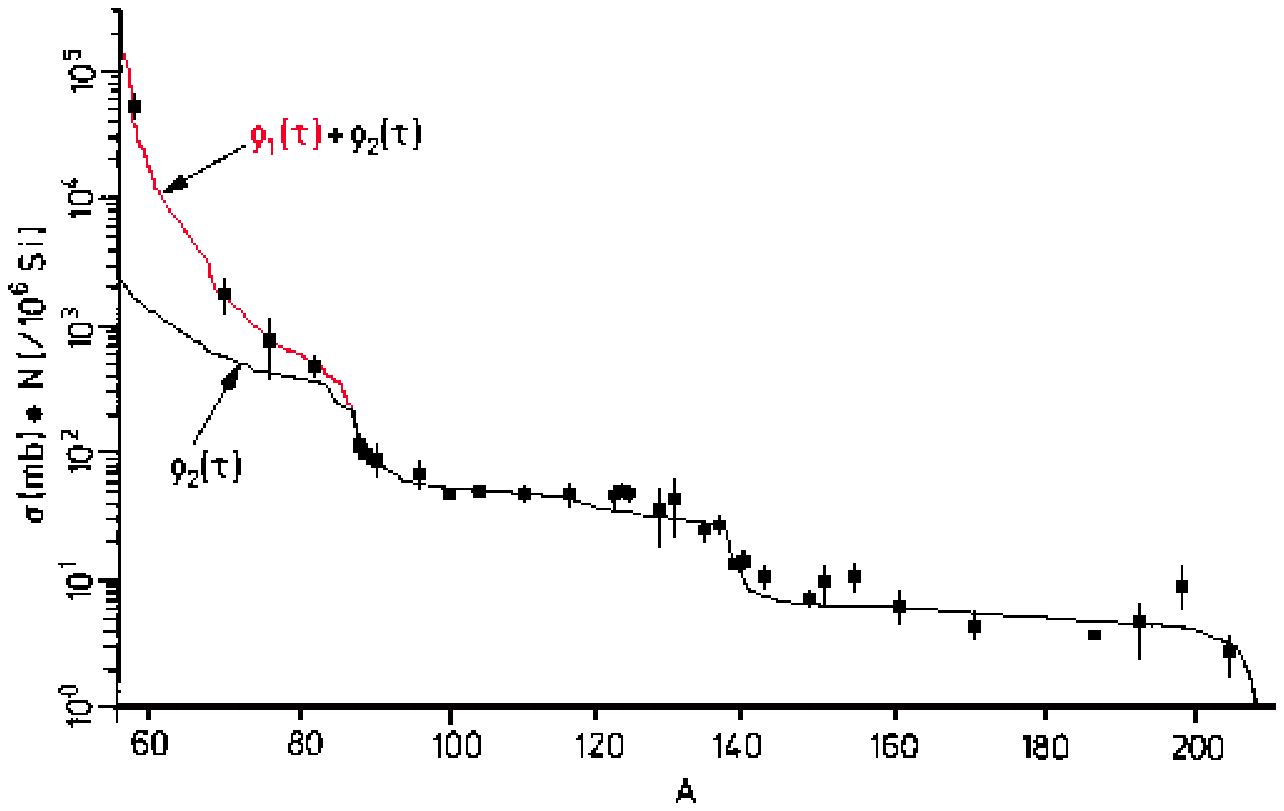}
\caption{Observed distribution of the product $\sigma_{A}\,N_A$ as a function of the mass number $A$ for the s-only nuclei. The neutron-capture cross section $\sigma$ is measured at 30 keV in millibarn, and the abundances are the solar ones (normalized to 10$^6$ Si atoms). The solid lines represent theoretical calculations for the single exponential distribution $\rho_2(\tau)$ (both panels) and for the sum of the exponential distributions $\rho_1(\tau)+\rho_2(\tau)$ (bottom panel). \citep[Figures adapted from][]{RR88}
\label{fig:component}}
\end{figure}

The process described by the set of coupled equations \ref{eq:fond_3} has the property to be ``self-regulating'' \citep[see also][]{Clay,RR88}, that is the tendency to reach a state of equilibrium where $\frac{dN_A(t)}{d\tau}\simeq 0 \rightarrow \sigma_{A-1}N_{A-1}\simeq \sigma_{A}N_A\simeq $ constant. For a true equilibrium condition in the s-process, the $\sigma\,N$ value should be strictly constant over the whole mass region from $^{56}$Fe up to $^{209}$Bi. However, looking at Fig.~\ref{fig:component}, it is easy to notice that such a condition is only satisfied locally in mass regions between magic neutron numbers and, for that reason, the condition is called the {\it local condition approximation}. 

The existence of a local equilibrium indicates that the observed distribution of $\sigma_{A}\,N_A$ can not be generated by a uniform exposure of iron-peak nuclei to a single neutron flux, but it is necessary a superposition of different exposures to various neutron fluxes.   
As a consequence, the quantity $\sigma_{A}\,N_A$ can be written as \citep[see e.g.][for details]{Clay}
$$\sigma_{A}N_A= \int_0^\infty\rho(\tau)\sigma N(\tau)\,d\tau,$$
where $\rho(\tau)$ represents the continuous distribution of multiple neutron exposures and is defined by the following relation
$$\rho(\tau)\varpropto\frac{1}{\tau_0}exp\left(-\frac{\tau}{\tau_0}\right),$$
in which $\tau_0$ is a parameter representing the mean neutron exposure.

It is found \citep[see e.g.][]{Clay,RR88,Kappeler99,Kappeler11} that at least two exponential neutron exposures, $\rho_1(\tau)$ and $\rho_2(\tau)$, are necessary to roughly reproduce the $\sigma_{A}\,N_A$ curve reported in Fig.~\ref{fig:component}. The first one, called weak component, has a relatively low mean neutron exposure ($\tau_{0,1}\simeq0.06$ neutrons millibarn$^{-1}$, with 1 neutron millibarn$^{-1}$ $\equiv$ 10$^{27}$ neutrons cm$^{-2}$) that allows for the synthesis of s-species in the  60~$\lesssim$~A~$\lesssim$~90 mass range (see also the red line in bottom panel of Fig.~\ref{fig:component}). The second one, called main component, has a higher mean neutron exposure ($\tau_{0,2}\simeq0.25$ neutrons millibarn$^{-1}$) and allows for the synthesis of the remaining s-nuclei.

\subsection{The s-process branchings}
\label{sec:branchings}

\begin{figure}[ht]
\centering
\includegraphics[width=12.0cm]{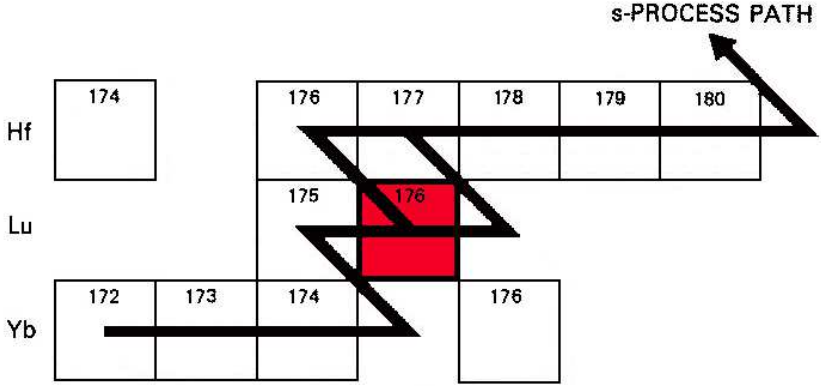}
\caption{A section of the chart of nuclides showing the s-process branching at the isotope $^{176}$Lu (highlighted in red). \citep[Figure adapted from][]{RR88}
\label{fig:branchings}}
\end{figure}

As above mentioned, the analytical approach discussed in the previous Sect.~\ref{sec:analytical_approach} is based on the assumption that the relation $\tau_{\beta} \gg \tau_{n\gamma}$ is valid for all the unstable nuclides involved in the nucleosynthesis process. However, the s-process path can actually encounter unstable nuclei for which the decay rate becomes comparable to the neutron capture rate (i.e.~$\tau_{\beta} \simeq \tau_{n\gamma}$). This leads to a splitting of the path, which is called s-process branching. As a consequence, a fraction of the s-process flow proceeds through neutron capture, while the other one goes through the $\beta$-decay (see Fig.~\ref{fig:branchings}). 

A comparison of abundances of nuclei reached at a branching starting at a nucleus with mass number $A$ can provide information on the physical conditions of the environment in which the s-process takes place, through the analysis of the so-called branching ratio $R$ defined as $$ R=\frac{1}{\tau_{\beta}\,N_n\,<\sigma v>_A}.$$ 

Indeed, the above mentioned ratio can be also deduced directly from the observations, considering that $R$ can be also written in terms of the observed values of $\sigma\,N$ as $R=\frac{(\sigma\,N)_{Z+1}}{(\sigma\,N)_{A+1}}$ or $R=\frac{(\sigma\,N)_{A-1}}{(\sigma\,N)_{A+1}} - 1$. As a consequence, if $\tau_{\beta}$ is temperature-independent and $<\sigma v>_A$ is known, the observed ratio $R$ can provide information about the s-process neutron density $N_n$. Instead, if $N_n$ is known, the ratio $R$ can become a sensitive s-process ``thermometer'' or ``barometer'', every time that the s-process branching involves reactions which critically depend on temperature or density \citep[see e.g.][for details]{RR88}.

\subsection{Actual sites of the s-process}
To have a more complete picture of the s-process it is necessary to individuate the astrophysical sites where the s-process can take place, keeping in mind that more than one s-process component is required in order to explain the observed solar distribution of s-nuclei abundances.

Current views on the subject suggest that the main and weak components of s-process correspond, in terms of stellar environments, to two distinct categories of stars in different evolutionary phases \citep[e.g.][]{Kappeler99}. 

In particular, the main component is associated with low-mass stars ($M_{ZAMS} \sim 1.5-3 M_\odot$) during their thermally pulsing asymptotic giant branch (TP-AGB) phase, when the H- and He-burning shells surrounding the degenerate stellar core are alternately activated. The predominant neutron source is the $^{13}$C($\alpha$,n)$^{16}$O reaction, but the $^{22}$Ne($\alpha$,n)$^{25}$Mg reaction can be also marginally activated, leading to a variation of some abundance ratios of nuclides belonging to s-process branchings \citep[for details on main component the interested reader is referred to the reviews of][and references therein]{Busso99,Kappeler99,Kappeler11}. 

The weak component occurs in massive stars ($M_{ZAMS} \gtrsim 13 M_{\odot}$) primarily during their core He-burning phase, and the most important neutron source is the $^{22}$Ne($\alpha$,n)$^{25}$Mg reaction. This reaction is efficiently activated only at the end of the core He-burning phase, when the temperature is $\gtrsim 2,5\times 10^8~K$. The available $^{22}$Ne is produced at the beginning of the core He-burning phase via the reaction sequence  $^{14}$N($\alpha$,$\gamma$)$^{18}$F($e^+\,\nu$)$^{18}$O($\alpha$,$\gamma$)$^{22}$Ne, where $^{14}$N  derives from the CNO cycle activated during the previous H-burning phase \citep[for details see e.g.][and references therein]{woosley2002}. 



In addition to these two components, other kinds of stars, such as massive AGB ($M_{ZAMS} \sim 4-7 M_\odot$) and super-AGB stars ending their life as NeO white dwarfs ($M_{ZAMS} \sim 7.5-10 M_\odot$; for details see e.g.~Fig.~1 in \citealt{pumo2009b}, but also \citealt{ps07} or \citealt{pumo07} and references therein), could also contribute to the nucleosynthesis of s-species, but this hypothesis still needs further investigation \citep[][and references therein]{pumo2009a}. Moreover, some studies \citep*[][]{gallino98,Busso99,lugaro03,gorielysiess2004} suggest the existence of a ``strong'' component, which occurs in low-metallicity stars of low-intermediate mass during the TP-AGB phase, and which is supposed to be responsible for the synthesis of ``massive'' (around $^{208}${Pb}) s-species. Furthermore, \citet[][]{travaglio04} propose the existence of an additional component referred to as lighter element primary s-process (LEPP), but its nature is still unclear and under debate \citep*[e.g.][and references therein]{tur09,pignatari10}.

\section{Production of s-nuclei in massive stars}
\label{sec:main}

\subsection{Sensitivity to stellar mass and metallicity}
\label{sec:sensitivity}

There is a wide consensus about the main characteristics of weak component of the s-process and, in particular, about its sensitivity to stellar mass and metallicity \citep[e.g.][]{Kappeler99}. 

\begin{figure}[ht]
\centering
\includegraphics[width=12.5cm]{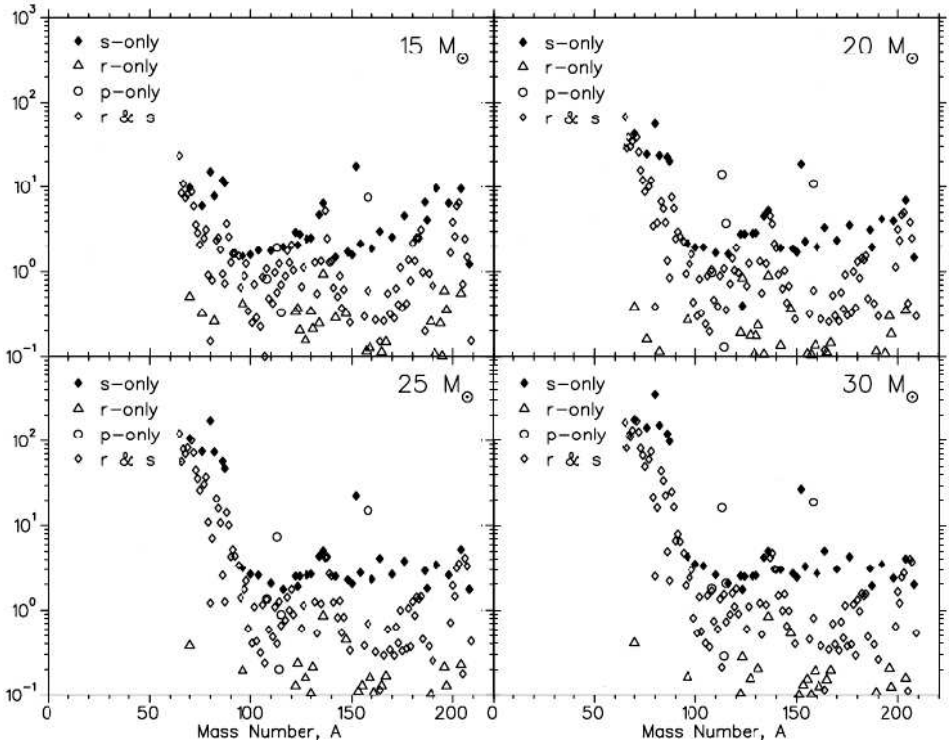}
\caption{Overproduction factor at the end of the core He-burning as function of the mass number $A$ for stellar models of different initial mass (namely, $M_{ZAMS}$~=~$15$, $20$, $25$, and $30$\Msun). All the stable trans-iron nuclei of the network used to simulate the s-process are reported. The primary nucleosynthesis production process of each nucleus is indicated by the symbol type. \citep[Figure adapted from][]{the07}
\label{fig:mass}}
\end{figure}

As for the dependence on the stellar mass, quantitative studies (see e.g. \citealt{prantzos90}; \citealt{kappeler94}; \citealt{rayethashimoto2000}; \citealt{the2000}, \citeyear{the07}) show that the s-process weak component efficiency\footnote{Usually the s-process efficiency is analyzed in terms of the following efficiency indicators (see e.g. \citealt{prantzos87}; \citealt{the2000}; \citealt{costa2006}; \citealt{pumo2006}, \citeyear{pumo10}):
\begin{itemize}
 \item[-] the average overproduction factor F$_0$ for the 6 s-only nuclei $^{70}$Ge, $^{76}$Se, $^{80}$Kr, $^{82}$Kr, $^{86}$Sr and $^{87}$Sr, given by 
 $$ F_0=\frac{1}{N_s}\sum_{i}F_i \mbox{\,~with\,~ } F_i=\frac{X_i}{X_{i,ini}},\,\,N_s=6$$ 
 where F$_i$ is the overproduction factor, X$_i$ is the mass fraction (averaged over the convective He-burning core) of s-only nucleus $i$ at the end of s-process, X$_{i,ini}$ is the initial mass fraction of the same nucleus, and N$_s$ is the number of the s-only nuclei within the mass range $60 \leq A \leq 87$;
 \item[-] the maximum mass number A$_{max}$ for which the species in the $60 \leq A \leq A_{max}$ mass range are overproduced by at least a factor of about 10;
 \item[-] the number of neutrons captured per initial $^{56}$Fe seed nucleus n$_c$ defined as 
 $$n_c=\sum_{A=57}^{209}(A-56)\frac{[Y_A-Y_A(0)]}{Y_{56}(0)}$$ 
 where $Y_{56}(0)$ is the initial number fraction of $^{56}$Fe, $Y_A$ is the final number fraction of the nucleus with mass number $A$, and $Y_A(0)$ is the initial one.
\end{itemize}
In addition to the previous parameters, the maximum convection zone mass extension and the duration of the nucleosynthesis event are also used for characterizing the efficiency of the s-process weak component during the core He-burning.
\label{note:efficiency}} decreases with decreasing initial stellar mass, and that the shape of the distribution of the overproduction factors as a function of the mass number essentially does not depend on the initial stellar mass value. This behavior is connected to the fact that the reaction $^{22}Ne(\alpha,~n)^{25}Mg$ becomes efficient only for $T\gtrsim 2,5\times 10^8~K$, so the production of s-nuclei is more and more efficient when the initial stellar mass is increased, because more massive models burn helium at a ``time averaged'' higher temperature; however the ratio of the overproduction factor F$_i$ of a given s-only nucleus $i$ to the average overproduction factor F$_0$ (see footnote \ref{note:efficiency} for details on F$_i$ and F$_0$) remains fairly constant irrespective of the stellar mass, so the shape of the distribution of the overproduction factors does not change when the initial stellar mass is increased (see also the behavior of the s-only nuclei distribution in Fig.~\ref{fig:mass}).

As for the effect of metallicity, the s-process weak component efficiency depends on the so-called {\it source/seed} ratio\footnote{Considering that the $^{22}$Ne is the main neutron provider during the core He-burning phase and neglecting all the heavier $^{56}$Fe nuclei, one obtains: $$ \mbox{source/seed} \simeq \mbox{$^{22}$Ne/$^{56}$Fe}.$$ This last quantity approximately corresponds to the $^{14}$N/$^{56}$Fe ratio at the end of the core H-burning phase which, in turn, is roughly equal to the O/$^{56}$Fe ratio at the ZAMS \citep[see][for details]{prantzos90,rayethashimoto2000}.} \citep[see e.g.][]{prantzos90,rayethashimoto2000}. If the {\it source/seed} ratio is constant with the metallicity Z, the efficiency is expected to increase when increasing the Z value, because the effect of the $^{16}$O primary poison becomes less important when the abundances of the {\it source} nuclei increase with Z. For a non-constant {\it source/seed} ratio that increases when decreasing Z, the efficiency (measured in terms of the number of neutrons captured per initial $^{56}$Fe seed nucleus n$_c$; see also footnote \ref{note:efficiency}) has a non-linear behavior with Z, which reflects the interplay between two opposite factors: from one hand, the aforementioned role of the $^{16}$O primary poison, which tends to decrease n$_c$ with decreasing Z, because its abundance remains the same independently of Z, so its relative importance increases as Z decreases; on the other hand the effect of the increased {\it source/seed} ratio, which tends to increase n$_c$ with decreasing Z, because the number of available neutrons per nucleus {\it seed} increases as Z decreases.

\subsection{Sources of uncertainties}
\label{sec:uncertainties}
Although the general features of the s-process weak component seem to be well established, there are still some open questions linked to the nuclear physics, the stellar evolution modeling, and the possible contribution to the s-nucleosynthesis from post-He-burning stellar evolutionary phases \citep[see e.g.][]{woosley2002,pumo2006,costa2006}.

The uncertainties due to nuclear physics are linked both with the reaction rates of reactions affecting the stellar structure evolution (as, for example, the triple-alpha, the $^{12}$C($\alpha$,$\gamma$)$^{16}$O and the $^{12}$C + $^{12}$C reactions) and with reaction rates on which the so-called ``neutron economy'' (i.e.~the balance between neutron emission and captures) is based. Many works \citep*[see e.g.][]{kappeler94,rayethashimoto2000,the2000,the07,hoffman2001,tur07,tur09,pignatari10,bennett10} have been devoted to analyze these uncertainties, showing that such uncertainties still affect significantly the s-process weak component efficiency. 

The contribution to the synthesis of s-nuclei during the post-core-He-burning evolutionary phases was also explored by many authors (see e.g. \citealt{arcoragi1991}; \citealt{raiteri1993}; \citealt{the2000}, \citeyear{the07}; \citealt{hoffman2001}; \citealt{raucher02}; \citealt{LC03}; \citealt{tur07}, \citeyear{tur09}), including in some cases the explosive burning (see e.g. \citealt{hoffman2001}; \citealt{raucher02}; \citealt{LC03}; \citealt{tur07}, \citeyear{tur09}). All these studies have shown that a significant production of s-nuclei in massive stars can continue during the post-core-He-burning evolutionary phases and that the abundance of s-nuclei ejected in the interstellar medium after the core-collapse supernova events can be substantially modified by the explosive nucleosynthesis. 

As for the impact of uncertainties owing to stellar evolution modeling, the determination of the size of the convective core and, more in general, of the mixing regions represents one of the major source of uncertainties still affecting the s-process weak component efficiency, as described in detail in the next Sect.~\ref{sec:overshooting}.

\section{Role of the convective overshooting}
\label{sec:overshooting}
The determination of the mixing regions and, in particular, of the size of the convective core can directly affect the efficiency of the s-process nucleosynthesis by influencing the chemical and local temperature stratification \citep[see e.g.][]{canuto97,MF94,DX08}, by determining the amount of stellar material which experiences neutron irradiation \citep[see e.g.][]{langer89}, and by giving rise to a variation of the s-process lifetime \citep[see e.g.][]{costa2006,pumo10}.

The convective core's extension of a star with a given initial mass and metallicity is determined in turn by a series of physical parameters such as the choice of the convective instability criterion (Schwarzschild's or Ledoux's criteria), the extra mixing processes induced by axial rotation and convective overshooting \citep[see e.g.][]{Chiosi92,woosley2002}.

A series of studies have been devoted to analyze the effects of these physical parameters on the evolution of massive stars (see e.g.~\citealt{MM97}, \citeyear{MM00}; \citealt{Heger00}; \citealt{woosley2002}; \citealt{Hirschi04}; \citealt{LC06}; \citealt{eleid09}) and to examine the corresponding impact on the s-process weak component \citep[see e.g.][]{langer89,pumo2006,costa2006,pignatari08}. 

\begin{figure}[ht]
\centering
\includegraphics[width=9.0cm]{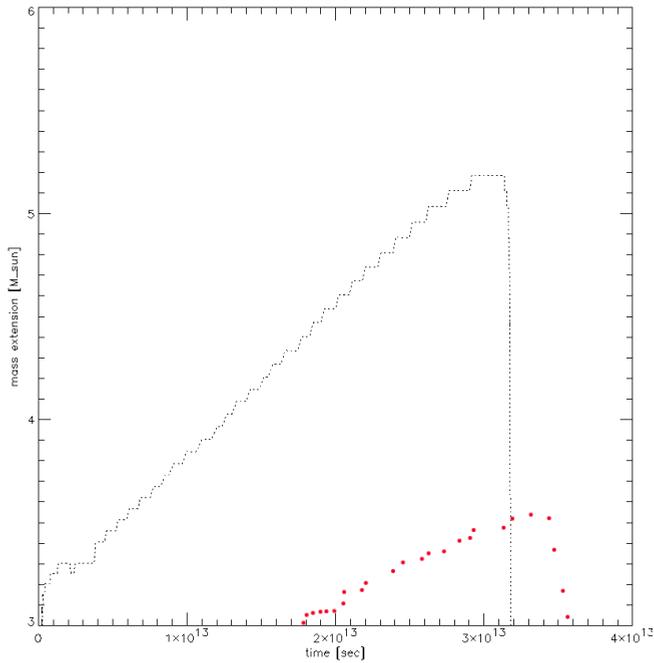}
\caption{Mass extension of the convection zone (in unit of $M_{\odot}$) during the core He-burning s-process as a function of time (measured from the core He-burning ignition) for a Z=0.005, M=20\Msun\, model with (dotted line) and without (red dots) convective overshooting. Details on the input physics used to calculate the models can be found in \citet[][]{pumo10}.
\label{fig:conv_zone}}
\end{figure}

As far as the convective overshooting is concerned, one finds that this extra mixing process leads to an increase of the convective core mass (see also Fig.~\ref{fig:conv_zone}) and to a variation of the chemical and temperature stratification that, in turn, tend to enhance the s-process weak component efficiency, by giving rise to an increase of the amount of material that experiences neutron irradiation and to a variation of the s-process lifetime \citep[see e.g.][for more details]{costa2006}.

Recent works \citep[][]{costa2006,pumo2006,pumo10} have been devoted to perform a comprehensive and quantitative study on the impact of the convective overshooting on the s-process, using a diffusive approach to model the convective overshooting \citep[for details see e.g.][and reference therein]{freytag96,herwing97}. The results show that models with overshooting give a higher s-process efficiency compared with ``no-overshooting'' models, with enhancements for the main s-process indicators $F_0$ and $n_c$ until a factor $\sim 6$ and $\sim 3$, respectively. 

\section{Final remarks}
\label{sec:remarks}
The results reported in Sect.~\ref{sec:overshooting} clearly show the high level of uncertainty (up to a factor $\sim 6$) in the modeling of the weak s-process component due to the current lack of a self-consistent theory describing mixing processes inside the stars, indicating that a detailed scrutiny of the impact of the stellar evolution modelling uncertainties on this component remain to be performed prior to giving a final conclusion on the the s-process weak component efficiency.

In particular, prior to giving a final conclusion on the possible contribution of post-He burning phases to the s-process yields from a quantitative point of view, some additional investigation taking into account stellar evolution uncertainties in addition to the nuclear physics ones should be performed. Moreover, this additional investigation may shed light on different open questions \citep[see also][]{pumo10} linked, for example, to the effective existence of the LEPP process and to the model for the p-process taking place in the core-collapse supernovae O-Ne layers \citep[see also][and references therein]{costa2006}, because the relevant s-nuclei are p-process seeds \citep[cf.~Sect.\ref{sec:nucleosynthesis} and see also e.g.][and references therein]{AG03}.

\section*{Acknowledgments}
M.L.P. acknowledges financial support from the Bonino-Pulejo Foundation and from the PRIN-INAF 2009 ``Supernovae Variety and Nucleosynthesis Yields''.

\end{document}